# Growth of transparent $Zn_{1-x}Sr_xO$ ($0.0 \leq x \leq 0.08$) films by facile wet chemical method: Effect of Sr doping on the structural, optical and sensing properties

Amit Kumar Rana[1], Rajasree Das[1,2], Yogendra Kumar[2], Somaditya Sen[1,2], Parasharam M. Shirage[*,1,2]

[1]Department of Physics, Indian Institute of Technology, Khandwa Road, Indore-452020 India.

[2]Centre of Materials Science and Engineering, Indian Institute of Technology, Khandwa Road, Indore-452020 India.

*Author for correspondence (E-mail: pmshirage@iiti.ac.in, paras.shirage@gmail.com)

**ABSTRACT**:

$Zn_{1-x}Sr_xO$ ($0.0 \leq x \leq 0.08$) nano-rods thin films are prepared using wet chemical technique on transparent flexible substrate. Effect of Sr-doping on structural and optical properties of ZnO is systematically investigated. SEM and TEM confirm the nano-rods like morphology with single crystalline nature of all the samples. Rietveld refinement of XRD shows the samples belongs to *P63mc* space group, furthermore, a gradual increment in lattice parameters and change in Zn/oxygen occupancy ratio is observed with Sr doping. SIMS and XPS confirm the desired elemental concentration in the nanostructures. XPS measurements shows that increase in Sr doping creates more oxygen associated defects, which is further supported by the photoluminescence spectra indicating the gradual change in Zn vacancy ($V_{zn}$) and oxygen interstitial ($O_{in}$) point defect intensities in the films. Near band edge emission peak shows to shift toward higher wavelength in the doped films. Pure ZnO film shows Raman peaks around $99(E_2^{low})$, $333(E_2^{high} - E_2^{low})$, 382 ($A_1$ (TO)), $438(E_2^{high})$ and 582($A_1$ (LO) +$E_1$ (TO)) cm$^{-1}$, whereas two additional defect driven vibrational modes (at 277 and 663 cm$^{-1}$) are appeared in the Sr-doped films. The sensing property of the ZnO is enhanced by Sr doping and imitates as a promising material for future toxic and flammable gas sensor applications as well as for opto-electronic devices.

# 1. INTRODUCTION

In recent few years, transparent ZnO thin film photoconductors have drawn the gigantic attention of the scientific community as they have great applications in environmental monitoring, large-area displays, and optical communications. Transparent semiconductors on flexible substrates provide multidimensional approach for technologies of the aforesaid applications and new opto-electronics devices emerging. Recently, few electronics companieshave showcased the world's first foldable active-matrix organic light-emitting diode (AMOLED) display. For potential application, ZnO and doped ZnO are promising candidates. So far Al, Co, Mn, *etc.* doping in ZnO are intensively carried out on various substrates. In the present research paper we demonstrate the facile wet chemical technique for the growth of least studied Sr-doped ZnOnano-material on flexible transparency substrate and provide the structural and optical properties measurements, which are important for the technological applications.

Zinc oxide (ZnO), a most common example of *n*-type oxide semiconductors, is well known for its wide band gap ($E_g$~ 3.37 eV at room temperature (RT)) and multifunctional properties (semiconducting, spintronic, pyro-electric and piezoelectric) [1-4]. Moreover, due to versatile applications of ZnO in the field of UV light emitter, transparent electronics, chemical sensors, spin electronics *etc.*, remarkable theoretical[5-7]and experimental[8-12]research workhas been done focused to its crystallographic structure, chemical compositions (defect mechanism) and physical properties. ZnOcrystallizes in the wurtzite structure and is stable up to a high temperature range (~2000ºC). Being rich with diverse structural and physical properties, nanostructures of ZnO are the building block of new generation nano-devices. Thus following an easy and cost effective route to develop 2D ZnO nanomaterial is fundamentally important. Focusing on the application point of view in nano-electronics, optoelectronics *etc.*, ZnOnanostructures have as much potential applications as carbon nanotubes or silicon nanowires, in addition it is widely used as potential substrate to grow GaN film(one of the blue light emitting diode material)[8].Doping with different ions and addition of impurity atoms is a well-acceptedapproach to tailor the electrical and optical properties in ZnO. Recent reports showroom temperature ferromagnetism (RTFM) in transition metal doped ZnObaseddiluted magnetic semiconductors (DMS) [13]. Interestingly, reports addressing the origin of this RTFM in DMSprovide conclusive proof that magnetic ion doping is not essential for FM behavior in ZnO[14], instead defects like oxygen vacancy ($V_O$) and Zn interstitial ($Zn_{in}$) plays important role here[15-16]. However, despite all this findings experimental study on defects and optical properties of ZnO is rather limited, particularly on $Zn_{1-x}Sr_xO$ compound. Detail understanding regarding defect driven luminescence and phonon vibrational modes is very necessary to develop high quality optoelectronic devices. $Zn_{1-x}Sr_xO$ shows promising feature to fabricate transparent electrodes in solar cells, love wave filter applications, Micro-electromechanical systems (MEMS), and ultrasonic oscillators[17]. Alkaline earth elements, especially those with large cationic radius, is reported to greatly influence both the varistor properties and microstructure. Water *et al.* reported that thermal treatment of Sr doped ZnO thin film in a reducing atmosphere modifies optical and electrical properties of the films[18]. Vijayan *et al.* showed enhancement in band gap and high gas sensitivity in the Sr doped ZnO films, deposited with a chemical bath deposition technique[19]. Binary ($Zn_{1-x}Ba_xO$) ceramics are susceptible to attack by moisture[20-21]. Investigations focused on developing sophisticated nanoscale materials like nanoparticles, nanotubes, nanowires, nanorods, nanosheets, nanoclusters, nanocones *etc.* are still desirable due to their importance in device fabrication and other practical applications such as promising photo catalytic activity (CdS and CdSe)[22], electro-catalytic activity[23], field emission properties ($MoS_2$ and Si doped AlN)[24]. So, Sr doped ZnO thin film is a very interesting field to explore as this is one of the least studied materials which is an important for technological applications.

In this study, we have developed a facile wet chemical technique to grow highly crystalline $Zn_{1-x}Sr_xO$ ($0.0 \leq x \leq 0.08$) thin films with nano-rods like surface morphology on flexible transparent substrates. We have successfully avoided high temperature, high vacuum, microwave assistance, auto-claves and other sophisticated ways to grow ZnO nano-materials. Here, a very detail investigation on structural and optical properties of Sr doped ZnO nano-rods are delivered. To the best of our knowledge, there is very less or no report available on SIMS, photoluminescence (PL) Raman spectroscopy and gas sensing study on $Zn_{1-x}Sr_xO$ ($0.0 \leq x \leq 0.08$) nanostructure.

## 2. EXPERIMENTAL TECHNIQUES

The $Zn_{1-x}Sr_xO$ ($0.0 \leq x \leq 0.08$) films were prepared on flexible transparency substrate (G-LAMI-ACE Product, A4 size, 100 micron thick transparency films) using wet chemical process from high purity nitrates of 100mM concentration. An aqueous solution of Zn nitrate hexahydrate ($Zn(NO_3)_2 \cdot 6H_2O$) and strontium nitrate ($Sr(NO_3)_2$) was prepared in double distilled water and stirrer for 30 min. Then aqueous $NH_3$ was added under constant stirring. A white precipitate was initially observed, which subsequently dissolved back in solution upon further addition of aqueous $NH_3$. The pH of the solution was maintained at ~12. After the pH was maintained for 30 min, carefully cleaned transparent substrates were dipped vertically and the solution was heated at 90°C for ~60 min. Finally, thin layer of material were coated on the substrates, removed from the solution, washed with distilled water and dried in air for overnight. All the transparent films were annealed at 100°C for 2 hrs.to get phase pure films. The crystallinity and morphology of the pure and doped ZnO ($Zn_{1-x}Sr_xO$ ($0.0 \leq x \leq 0.08$)) films were investigated by x-ray diffraction (XRD) using Bruker D8 Advance X-ray diffractometer (Cu $K_\alpha$ source). Field emission scanning electron microscope (FESEM) was carried out by using (Supra55 Zeiss). HRTEM images were taken by HRTEM, JEOL JEM 2100 and JEOL 3010 with a UHR pole piece. Secondary ion mass spectrometry (SIMS) was performed on the Sr doped ZnO films in the Hiden Analytical SIMS Workstation (Primary ion $O^{2+}$, Energy 5keV, Beam Current 400nA). Measurement was performed in presence of electron flood gun (500eV) to avoid charging at a base pressure of $8\times10^{-10}$ mbar and operating pressure $8\times10^{-8}$ mbar. Intensity of the sputtered element depends on many factors like sputtering yield, ionization probability of the individual elements *etc.*[25] and the elemental intensity ratio for a simple compound can be expressed as[26],

$$\frac{C_A}{C_B} = k \frac{I_A}{I_B}$$

Where, the proportionality constant *k* is the relative sensitivity factor (RSF) for compositional analysis, $I_A$ and $I_B$ are the intensity of A and B element respectively, and $C_A$ and $C_B$ are the surface concentration of A and B element, respectively, in the compound. In case of the Sr doped ZnO films, as the neighbor of the element (Sr or Zn) is similar and same experimental environment is maintained for all the films, the RSF value is considered to be identical for all samples.

Chemical composition and valance state of films wereexamined by X-ray photoelectron spectroscopy (XPS) performed at angle-integrated photoemission spectroscopy (AIPES) beam line on Indus-1 synchrotron radiation source at Raja Ramanna Centre for Advanced Technology (RRCAT), Indore, India. Room temperature PL spectroscopic measurements of the as synthesized films were conducted using a spectrofluorometer (Horiba JobinYvon, Fluorolog-3) having Xe lamp source with an excitation wavelength of 330 nm. Micro Raman scattering measurement were recorded using Labram-HR 800 spectrometer equipped with excitation radiation at wavelength of 488 nm from an argon ion laser at a spectral resolution of about 1 cm$^{-1}$. The sensing properties are measured by usingprogrammable Keithley source meter 2401, the sensing response in air or testing gas measured by monitoring current.

## 3. Results and Discussion

*ZnONanorods Growth Mechanism:*

The mechanism responsible for the growth of ZnOnanomaterialis Zn(OH)$_2$ or Zn(OH)$_4^{2-}$ as the growth units[27-29], follows the reactions from (1)-(6). The important parameters which are responsible for the growth of less than 200 nm diameter size nano-rods are; pH ~12, temperature ~ 90°C anddeposition time~60 min. These factors play crucial role for shape and size control of the nano-rods. For doped sample strontium nitrate is added with nominal composition from 0-8%, which follows the same mechanism as for ZnO.

$$Zn(NO_3)_2 + 2H_2O \xrightleftharpoons{NH_3,\ 90\ °C} Zn(OH)_2 + 2HNO_3 \quad \text{----------- (1)}$$

$$Zn(OH)_2 \leftrightarrow Zn^{2+} + 2OH^- \quad \text{---------- (2)}$$

$$Zn^{2+} + 4NH_3 \leftrightarrow Zn(NH_3)_4^{2+} \quad \text{----------- (3)}$$

$$Zn^{2+} + 4OH^- \leftrightarrow Zn(OH)_4^{2-} \quad \text{----------- (4)}$$

$$Zn(OH)_4^{2-} \leftrightarrow ZnO + H_2O + 2OH^- \quad \text{----------- (5)}$$

$$ZnO + 2OH^- \leftrightarrow Zn(OH)_2 \leftrightarrow ZnO + 2H_2O \quad \text{----------- (6)}$$

These reactionsare responsible for the growth of ZnO nano-rods thin film on the flexible transparent substrates like "trasparency", which showed nearly 95% of transmittance in the visible region (Supplementary Fig. S1) and after deposion the transmittance of pure and Sr 8% doped ZnOis nearly 80% and more in the visible region. The transmission sharply falls in the UV region due to the fundamental absorption.

Figure 1 represents the typical SEM image of (*a*) undoped and (*b*) 8 % Sr doped ZnO on transparancy sheets. Surface of all the films shows rod like morphology. The average diameter of the rods are ~ 100 nm for pure ZnO and almost double for the Sr-doped ZnO (~200 nm). The increase in diameter of the rods may be related to Sr-doping effect as Sr doping might enhances reaction process in the solution[30]. The length of the rods ranges for few micrometers.

The lattice structures and size of ZnOnanorods were characterized by HRTEM and Selective Area Electron Diffraction (SAED). Fig. 2(a-f) represents the TEM and HRTEM images of $Zn_{1-x}Sr_xO (0.0 \leq x \leq 0.08)$. The HRTEM images of the undoped and Sr doped ZnO (Fig.2) confirms the formation of rod like structure with diameter fluctuating from ~150-250 nm. All the images shows extended well developed lattice fringes throughout the highly crystalline ZnOnanorods, with a lattice spacing of about 0.26 nm in case of pure ZnO, which corresponds to the interspacing of (002) reflection planes. Whereas, the fringes of higher doped samples shows lattice spacing of about 0.28 nm and 0.24 nm which corresponds to the (100) and (101) reflection planes of ZnO hexagonal structure for Sr 8% doped samples. SAED pattern of ZnOnano-rod suggests single crystalline nature of the nano-rods (inset of Fig. 3(a)). To check the doping of the Sr in ZnO, the elemental color mapping TEM was carried out on 8% Sr doped ZnO. Fig 3 shows TEM image of $Zn_{0.92}Sr_{0.08}O$ nano-rods selected for elemental color mapping. Fig. 3(a), (b) and (c) represents elemental color mapping of Zn, Sr and O, respectively. Elemental mapping using TEM confirms the homogeneous distribution of Sr in the 8 % Sr doped ZnOnano-rods.

XRD patterns of $Zn_{1-x}Sr_xO$ $(0.0 \leq x \leq 0.08)$ thin films are shown in Fig. 4. Peak position of the filmsreveals that all peaks are associated tothe characteristic peaks of the hexagonal wurtzite structure with space group *P63mc*. No trace of impurity is found in $Zn_{1-x}Sr_xO$ $(0.0 \leq x \leq 0.08)$ films, which indicates the growth of high purity films. Reitvelt refinement is carried out to obtain detailed information of the crystal structure and lattice parameter of the $Zn_{1-x}Sr_xO$ $(0.0 \leq x \leq 0.08)$ films. **Table I** summarizes parameters obtained from the Reitvelt refinement. Increase in doping concentration shows peak shifttowards lower theta, which is due to the addition of higher radii $Sr^{2+}$(0.118 nm) ion in the lattice, replacing $Zn^{2+}$(0.074 nm). Fig. 5(a) shows change increase in the lattice parameter (*a* and*c*) with doping concentration which follows the Vegards Law. Existence of Zn and oxygen related defects are a well-known factor, our analysis shows a good agreement with it too. **Table I** shows ratio of Zn/oxygen occupancy is less than 1 in the ZnO lattice structure and Sr doping decreases this ratio further, which supports the mechanism of defect formation by doping[31].

SIMS is one of the most reliable analytical techniques to determine elemental concentrations in the compound[32]. Fig.6 shows that SIMS intensity signals of Sr in all the doped films remain almost constant over the entire period of sputtering time, indicating homogeneity in composition throughout the films. A careful observation of the data shows that the change in $C_{Sr}/RSF_{Sr}$ ratio is very systematic with increase in doping concentration in the films which confirms the successful doping of Sr in the ZnO matrix further.

Figure 7(*a*) and (*b*) shows high-resolution XPS spectra of asymmetric O 1s core peak fitted with two Gaussian components, of pure ZnO and with $x = 0.08$ ($Zn_{1-x}Sr_xO$) thin films, respectively. Lower binding energy (LBE) peak at 529.3 and 529.6 eV for $x= 0$ and $x = 0.08$, respectively, arises from $O^{2-}$ ions in a normal ZnO wurtzite lattice structure which forms Zn-O bond[33]. Whereas, higher binding energy (HBE) peak at 531.4 and 531.3eV for $x= 0$ and $x = 0.08$, respectively, is attributed to Zn-OH bond due to hydroxyl group[34], or other radicals on the film surface such as CO or $CO_2$ *etc*[33]. However, some contribution to this HBE peak of the O 1s spectrum can also be related to the presence of oxygen deficiency within the ZnO matrix[35-36]. In our case the changes in the intensity ratio of the LBE and HBE oxygen peak with Sr doping is worth noticing. As mentioned before, Zn/Oxygen occupancy decreases slightly **(Table I)** in the doped thin films indicating a large concentration of oxygen and Zn vacancies. Therefore, increase in the HBE peak intensity in the $x = 0.08$ film can be explained as, in the Sr doped films more 1s oxygen electrons are loosely bonded to lattice compared to the pure ZnO film.

The oxidation state of Zn can be determined by analyzing the Zn 2p spectra of the films, shown in Fig. 7 (*c*). Both the films possess characteristic doublet peak of Zn $2p_{3/2}$ and Zn $2p_{1/2}$ at 1021.5 and 1044.6 eV for ZnO and at 1021.2 and 1044.2 eV for $Zn_{0.92}Sr_{0.08}O$, respectively, along with their corresponding satellite peaks. The obtained binding energy (BE) difference between Zn $2p_{3/2}$ and $2p_{1/2}$ (23.1 and 23 eV for undoped and doped films, respectively) peaks are in good agreement with the reported value for $Zn^{2+}$ oxidation state[16, 37]. Fig.7(*d*) shows high-resolution XPS spectra of Sr 3p region of $Sr_{0.08}ZnO$ thin film. It reveals two distinct peaks at 269.3 and 279.4eV correspond to Sr $3p_{3/2}$ and $3p_{1/2}$ states, respectively[38, 39], having a BE difference of 10.1 eV. BE separation of the doublet 3p peaks for Sr-O bond is 10.1 eV whereas, for metallic Sr it is 10.4 eV[40]. Second peak at 274.1 eV in Fig. 7 (*d*) is the shake up or satellite peak of C1s. The chemical composition of the film is calculated according to the formula as follows[41]:

$$X(\%) = \frac{A_X/S_X}{\sum_{i=1}^{N} A_X/S_X}$$

Where $A_X$ ($A_i$) and $S_X$ ($S_i$) indicated the peak area and sensitivity factor of elements, $x(i)$ and $N$ is total the number elements. The results of the chemical composition, based on the intensities of the O, Zn and Sr lines are given in Table 2. XPS spectra of the Sr 3p region suggest successful incorporation of Sr in ZnO matrix replacing Zn and that Sr is in a 2+ oxidation state in the doped films

Recently, PL spectra of ZnO are investigated theoretically and experimentally by many research groups in some extent to understand the complex defect physics of ZnO structure. In ZnO, shallow Zn interstitials ($Zn_{in}$) and Oxygen vacancy ($V_O$) is responsible for the main visible emission, as in Zn-rich or O-rich conditions, $Zn_{in}$ and $V_O$ have low formation enthalpies[42]. Whereas, the existence of other intrinsic point defects and complexes (such as Zn vacancy ($V_{Zn}$), oxygen interstitials ($O_{in}$), antisite oxygen ($O_{Zn}$), complex of $V_O$ and $Zn_{in}$ ($V_O Zn_{in}$) and complex of $V_{Zn}$ and $Zn_{in}$ ($V_{Zn} Zn_{in}$) in ZnO are also reported previously. Van Dijken proposed that intensity of these defect emissions, especially exciton emission, strongly depends on the size of ZnO nanostructures[43]. This observation makes it more sensible about the different emission peaks and related defects observed by different research groups such as, Behera *et al.*[44] observed six different defect states consisting of emission from conduction band and/or defect donor sites to valence band and/or deep acceptor defects sites, Ghosh *et al.*[16] showed existence of seven emission peak in doped ZnO nanowires. Theoretical analysis carried out by Xu *et al.*[45] showed seven different defect states in ZnO, whereas, Zhang *et al.*[42] observed only five donor and acceptor type defect levels by local density approximation (LDA) theorem.

Figure 8(a) represents room temperature (RT) normalized PL spectrum and Gaussian fitted curves of ZnO thin film and Fig. 8 (b) shows the schematic band diagram of all the emissions assembled from the PL data. The typical wide emission spectrum of ZnO extending from near band edge (NBE) to green emission can be well resolved into 10 peaks at 378 (3.29 eV), 388 (3.21 eV), 404 (3.07 eV), 436 (2.85 eV), 450 (2.76 eV), 467 (2.66 eV), 480 (2.59 eV), 489 (2.54 eV), 499 (2.49 eV) and 562nm (2.20 eV). Existence of all these well resolved peaks is related to so many defects level in ZnO film. First peak ($E_1$) located in the UV region (378 nm) corresponds to the NBE in ZnO film, arises due to the free exciton (FX) recombination process via exciton–exciton collision[46]. Second emission peak ($E_2$) located at 388 nm with an energy difference of 77 meV with $E_1$ is reported to be attributed to the first longitudinal optical (LO) phonon replica of FX[47]. The emission peak $E_3$ located at 404 nm, originates from the singly ionized $V_{Zn}$[46]. Whereas, the violet emission peak at 436 nm ($E_4$) is originated due to electron transition between valance band maxima (VBM) and shallow donor level of the $Zn_{in}$[46, 47]. Low intensity peak at 450 nm ($E_5$) and 467 nm ($E_6$) is reported due to the carrier transition from exciton level to the $O_{in}$ level and $Zn_{in}$ to acceptor $V_{Zn}$, respectively[40]. Another low intensity blue peak at 480 nm ($E_7$), which is might be originated from conduction band minima to $V_O$[29].

Green emission or deep band emission (DBE) region in ZnO films need special attention as various interpretations can be found regarding the origin of PL band here. Børseth $et$ $al.$[48] reported the energy value of optical signal of $V_O$ and $V_{Zn}$ based on the data of annealing of samples in different atmosphere. Following the analysis and results reported by Leiter $et$ $al.$[49] and Børseth $et$ $al.$, green emission peak $E_8$, at 489 nm can be attributed to the $V_O$ defects. Similarly, green emission peak $E_9$ situated at 499 nm may be correlated to the defects result of deep level or trap-state emission[50]. Lin $et.al.$[51] theoretically showed that green emissions with peak at 2.38 eV can be assigned to the antisite oxygen ($O_{Zn}$). Yellow luminescence band peak $E_{10}$ at 564 nm is generally identified as dopant defect in wet chemical process grown ZnO[48] or even the interstitial oxygen ($O_{in}$) vacancy defect[50]. As in our case, structural measurements detect no trace of elemental impurity in the films, $E_{10}$ is attributed to the electronics transition from the bottom of the conduction band to $O_{in}$, which is close to the reported value (573 nm) by Liu $et$ $al.$[52].

Figure 9 shows clear change of the PL peak positions with doping concentration of alkaline metal ion Sr in ZnO film. **Table III** shows deconvolution of PL spectra of intrinsic and Sr doped films. Inset of Fig. 9 show a noticeable increase in relative intensity of $V_{Zn}$ ($E_3$ and $E_6$) and $O_{in}$ ($E_5$) related peak is observed with doping indicating that doping of Sr replaces Zn and creates more defects in the matrix. Lin $et.al.$[51] also mentioned similar phenomenon in their finding. However, it is worth to mention here that with increase in Sr doping emission intensity of VBM to $Zn_{in}$ ($E_4$) decreases consistently. Therefore, it is evident that relative intensity of this violet-blue luminescence peak decreases sharply due to incorporation of different Sr-defects such as Sr substitution ($Sr_{Zn}$) and/or Sr interstitials ($Sr_{in}$) defects in the lattice. Green emission peak intensity ($E_8$ and $E_{10}$) remains same in all samples as all were annealed in air atmosphere. This observation is inconsistent with the previous report on ZnO NPs, which shows that the incorporation of alkali metal ions increase the green band emission and yellow band emissions of ZnO[50]. Whereas, very small decrease in E$_9$ peak intensity is noticed. All the above mentioned changes in the PL spectrum further confirm Sr substitution in ZnO films, which can be safely concluded from the SIMS experiment too.

Group theory analysis of wurtzite ZnO structure, the total number of atoms per unit cell is s = 4, space group *P63mc*, at the Brillouin zone center (**q**=0) shows that phonon dispersion of ZnO has 12 branches consists of polar modes ($A_1$ and $E_1$), two non-polar modes ($2E_2$) and two silent ($2B_1$) Raman modes[53]. Among these, $A_1$ and $E_1$ polar modes are divided into 2 active phonon modes (transverse optical (TO) and longitudinal optical (LO)), $E_2$ is divided into $E_2^{low}$ and $E_2^{high}$ active modes and the $B_1$ branches are Raman inactive modes.

Figure 10 represents the RT Raman spectra of undoped and Sr doped films ranging from wavenumber 90 to 700 cm$^{-1}$. **Table IV** summarizes the Phonon mode frequencies (in units of cm$^{-1}$) of wurtzite Zn$_x$Sr$_{1-x}$O ($0.0 \leq x \leq 0.08$) films. Raman spectra of ($0.0 \leq x \leq 0.08$) films shows similar feature but the subtle aspects like intensity and half-width of the peaks shift with Sr doping can be evidenced, which is related to Raman inelastic scattering. Intensity of the peak at 438.2 cm$^{-1}$ (undoped ZnO), which is the characteristic mode for the wurtzite ZnO structure ($E_2^{high}$), decreases with doping while the peak position shifts very little (438.9 cm$^{-1}$) in $x = 0.08$ film (Inset of Fig.10). All the samples show Raman peaks around (P1) 99, (P3) 333, (P4) 382, (P5) 438 and (P6) 582 cm$^{-1}$. Peaks around 99 and 438 cm$^{-1}$, the most prominent peak of all the samples, are assigned to the $E_2^{low}$ and $E_2^{high}$ modes, respectively[29]. Modes near 333 and 382 cm$^{-1}$ is commonly assigned to ($E_2^{high} - E_2^{low}$) which is a 2$^{nd}$ order mode due to multi-phonon process and A$_1$ (TO), respectively[54]. According to the earlier reports two LO modes of ZnO is situated at 575 (A$_1$) and 587 (E$_1$) cm$^{-1}$[11]. Defect states in ZnO (V$_O$, Zn$_{in}$ or free carriers) contributes to the LO mode intensity of E$_1$ symmetry[12]. Mn and Co-doped ZnO shows significant change in E$_1$ (LO) mode due to the modification in intrinsic host lattice defects with doping[54, 55]. Whereas, Bundesmann *et al.*[11] and Kaschner *et al.*[56] have observed a peak at 583 and 582 cm$^{-1}$, respectively, only in the doped films and assigned it as additional peak (APs) due to defects. PL result of Sr doped and undoped film shows that concentration of V$_O$ and Zn$_{in}$ defect state is not increased with doping and all the samples shows a broad Raman scattering peak centered around 582 cm$^{-1}$, as a consequence this peak can be assigned to the combination of both LO vibrational modes of ZnO. An insignificant shift is demonstrated in the observed peak positions compared to the reported values as the phonon mode frequencies varies depending on the size, morphology and preparation techniques of the samples[44].

Apart from these five major peaks, doped samples show two more small intensity peaks (P2~277 and P7 ~663 cm$^{-1}$). Origin of these two peaks is not very clear yet as the earlier reports claims various reason of occurrence of these peaks. Kaschner *et al.*[56] correlated phonon vibration modes at 275 cm$^{-1}$ with N incorporation in ZnO structure, claiming that intensity of this peak increases linearly with N doping concentration. Meanwhile, Bundesmann *et al.*[11]observed similar peak (277 cm$^{-1}$) in Fe, Sb, Al, Ga, and Li doped ZnO films and reported it as a result of lattice defects. Following these literatures, APs at 277 cm$^{-1}$ in Sr doped ZnO film, which is more prominent in the $x$ =0.08 doped film, can be assigned to the defects created in the ZnO matrix due to doping. Last broad peak at 663 cm$^{-1}$ is also a reason of major debate. Serrano *et al.*[6] showed using two-phonon DOS calculation that intrinsic mode of ZnO at 650 cm$^{-1}$ arises from two phonon process TA+LO, while, Wang *et al.*[10] observed a peak at 663cm$^{-1}$ in Mn doped ZnO nanoparticles and reported it due to combined effect of $A_1$ (LO) + $E_2^{low}$ and existence of $Zn_2MnO_4$ precipitates. Meanwhile, Yang *et al.*[57] assigned this peak (at 660 cm$^{-1}$) to $V_O$ related intrinsic mode of ZnO. In our case, as both pure and doped ZnO films show $V_O$ related PL peak with an almost same intensity and Raman spectra of pure ZnO does not have any peak in this region, low intensity peak at 663cm$^{-1}$ can be assigned to merelybecause of the presence of $Sr^{2+}$ in ZnO lattice.

Nanomaterials are promising materials for the gas sensors which attracted significant interest due to considerable sensitivity and shorten response time. ZnO is one of the potential candidates among the nano-materials sensors[58].The *I-V* characteristic of ZnO and $Zn_{0.92}Sr_{0.08}O$ nanorodsstudied in ethanol and ammonia environment at room temperatureare shown in Fig. 11(*a*) and 11(*b*), respectively. Sr doped samples are more sensitive to ethanol and ammonia compared to the undoped ZnO nanorods. Whereas, Srdopednano-rods has better sensing property for ammonia than ethanol and air (Fig. 11(*c*)). It's because of improvement of the surface catalytic of doped samples with higher surface to volume ratio and also increase in oxygen related defected conform by XPS, which encourages the combustion of ammonia with the chemisorbed oxygen ions and increase it sensitivity as compare to ethanol[59, 60].The present results have been compared with other available reports of ZnO thin films[61, 62,63] and are given in supplementary Table S2. Figure 12 shows the transient resistance response of doped ZnO for 10 ppm concentrations of ammonia environment. The room temperature response of nanostructured for 10 ppm of ammonia was found to be high (Response $(S)$ = 138) with faster response and recovery times of 25 and 12 s respectively. These results indicating that Sr doped ZnO materials is promising materials for the ammonia sensing applications.

The generalized well-known mechanism known for the ZnO sensors is based on adsorption of the oxygen onto the surface which results in significant variation in the electrical resistance[64]. The formation of oxygen adsorbates ($O^{2-}$ or $O^-$) results in an electron-depleted surface layer due to electron transfer from the conduction band (CB) of the ZnO to oxygen, resulting in a decrease in electrical conductivity[65, 66]. When sensors are exposed to reducing gases, they can catalytically react with the oxygen adsorbates, freeing electron to return to the ZnO surface and increasing the sensors conductivity, thus, gases are detected. In the present investigation, results clearly show that the sensing performance of Sr-doped ZnO nano-rods is more preferable than those of ZnO nano-rods.

## 4. Conclusions

In summary, $Zn_{1-x}Sr_xO$ ( $0.0 \leq x \leq 0.08$ ) nano-rods thin films are successfully synthesized by facile wet chemical process on flexible transparency substrate, for the first time. SEM and TEM results show formation of nano-rods with diameter ~100-250 nm depending on Sr doping concentration. TEM with SAED provided important information that nano-rods are highly single crystalline in nature and while HRTEM results suggests orientation of rods along (00$l$) planes for 0-4% Sr doped samples and which then deviate for higher doping levels. Elemental color mapping by TEM and SIMS data indicates that Sr is indeed doped in ZnO. The systematic variation of the Sr-doping levels and Sr homogeneity in ZnO also confirmed and represented from SIMS experiment. Reitvelt refinement of the XRD data shows all the films crystallizes into wurtzite structure, corresponds to the space group *P63mc*, with a systematic increase in the lattice parameters with Sr-doping. Room-temperature XPS analysis provide strong evidence regarding creation of oxygen related defects in the doped films, while the valance state of Zn remains same in undoped and doped films. This finding is further explained qualitatively based on the PL data. Oxygen vacancy ($V_O$) concentration in all the films is almost same; while peak related to Zn vacancy ($V_{Zn}$) and oxygen interstitial ($O_{in}$) shows an obvious enhancement. Raman spectra of all the films shows characteristic phonon dispersion peak of wurtzite ZnO, while two additional peaks (APs) are observed in the doped films only, which provide the hints of doping dependent changes. These APs around 277 and 663cm$^{-1}$ are assigned to the intrinsic defects created in the ZnO matrix due to Sr doping. These optical observations may help to get further insight regarding development of better ZnO based optoelectronic devices. The sensing property of the ZnO is enhanced by Sr doping and reflects promising material for future sensor applications.

## Acknowledgments

This work was supported by the Department of Science and Technology (SERB-DST), India by awarding a prestigious 'Ramanujan Fellowship' (SR/S2/RJN-121/2012) to the PMS. PMS is thankful to Prof. Pradeep Mathur, Director, IIT Indore, for encouraging the research and providing the necessary facilities. We are also thankful for SIC lab of IIT Indore for providing necessary characterization facilities. We are thankful to Mukul Gupta, Vasant Sathe, D. M. Phase and Ram Janay Choudhary, IUC-DAE Consortium for Scientific Research, Indore, for their help to doSecondary ion mass spectrometry (SIMS), Raman measurement, SEM and X-ray photoelectron spectroscopy (XPS) of the samples respectively. Authors also acknowledge Professor T. Pradeep, and DST Unit of Nanoscience at IIT Madras and Mr. Samik Roy Moulik form Icon Analytical Equipment Pvt. Ltd. for providing TEM Images.

**Figures Captions:**

Figure 1. Field Emission Scanning Electron Microscope (SEM) of (a) ZnO and (b) Sr-doped ZnO thin films.

Figure 2(a-f). TEM and HRTEM images of $Zn_{1-x}Sr_xO$ ($0.0 \leq x \leq 0.08$) of (a-b) $x=0$, (c-d) $x=0.04$ and (e-f) $x=0.08$ respectively.

Figure 3 (Color online) Elemental mapping of 8% Sr doped ZnO nanorod, (a) Zn, (b) O and (c) Sr. Inset of (a) shows Selective Area Electron Diffraction (SAED) of ZnO nano-rod.

Figure 4. (Color online) XRD patterns of (a) undoped ZnO, (b) $Sr_{0.02}ZnO$, (c) $Sr_{0.04}ZnO$, (d) $Sr_{0.06}ZnO$ and (e) $Sr_{0.08}ZnO$ thin films.

Figure 5. (Color online) The variation of ZnO lattice parameters, $a$ and $c$ as a function of Sr doping concentration. Inset of the figure shows corresponding change in lattice volume with Sr doping concentration.

Figure 6 Secondary Ion Mass Spectroscopy (SIMS) of $Zn_{1-x}Sr_xO$ ($0.0 \leq x \leq 0.08$) nano-rods films with $x = 0, 0.02, 0.04, 0.08$.

Figure 7. (Color online) XPS spectra of the (a) O1s peak in undoped ZnO film, (b) O1s peak in $Zn_{0.92}Sr_{0.08}O$ film (c) Zn 2p peak in ZnO and $Zn_{0.92}Sr_{0.08}O$ film and (d) Sr 3p peak in $Zn_{0.92}Sr_{0.08}O$ film.

Figure 8. (Color online) (a) Room temperature PL spectra of undoped ZnO thin films. The "■" shows experimental data and solid green and red lines are Gaussian fitting of individual peaks and sum of all peaks, respectively (b) Schematic band diagram for ZnO thin film.

Figure 9. (Color online) Experimental data for room temperature PL spectra of $Zn_{1-x}Sr_xO$ ($0.0 \leq x \leq 0.08$) thin films. Inset shows the change in relative intensity of $E_3$ ($V_{Zn}$ related peak) and $E_5$ ($O_{in}$ related peak) peak with Sr doping concentration in the films.

Figure 10. Room temperature Raman spectra of $Zn_{1-x}Sr_xO$ ($0.0 \leq x \leq 0.08$) thin films. Inset shows the relative shift of $E_2^{high}$ mode with respect to Sr concentration in the films.

Figure 11. *I-V* characteristic of ZnO and $Sr_{0.08}Zn_{0.92}O$ in (a) ethanol and (b) ammonia. (c) Shows *I-V* characteristic of $Sr_{0.08}Zn_{0.92}O$ nano-rods in air, ethanol and ammonia environment.

Figure 12. (a) Resistance variation of the $Sr_{0.08}Zn_{0.92}O$ for 10 ppm of $NH_3$ at room temperature (b) Transient resistance response of ZnO thin film towards 10 ppm of $NH_3$.

**Table Captions:**

**Table I**. *Refined structural parameters for $Zn_{1-x}Sr_xO$ ($0.0 \leq x \leq 0.08$) thin film. (Error in $10^{-4}$ order in refined parameters)*

**Table II.** *Chemical composition of Zinc, Oxygen and Strontium measured by XPS.*

**Table III**. *Peak position and relative peak intensity of deconvoluted PL spectra of $Zn_{1-x}Sr_xO$ ($0.0 \leq x \leq 0.08$) thin films. Peak position (nm) and relative intensity (%) have the errors $\pm 0.5$ and $\pm 1$ respectively.*

**Table IV.** *Phonon mode frequencies (in units of cm-1) of wurtzite $Zn_{1-x}Sr_xO$ ($0.0 \leq x \leq 0.08$) thin films.*

*Table I*

| Sample | Lattice parameters (Å) | Atoms | | Positions | | | Occ. Ratio (Zn/oxygen) | Volume (Å$^3$) |
|---|---|---|---|---|---|---|---|---|
| | | | | X | Y | Z | | |
| ZnO | $a$=3.2434±2 | Zn | 2a | 0 | 0 | 0.8555±3 | 0.503 | 47.332±2 |
| | $c$=5.1952±1 | O | 2b | 0.3333 | 0.6667 | 0.8800±2 | | |
| Zn$_{0.98}$Sr$_{0.02}$O | $a$=3.2450±2 | Zn/Sr | 2a | 0 | 0 | 0.3950±1 | 0.417 | 47.402±6 |
| | $c$=5.1980±2 | O | 2b | 0.3333 | 0.6667 | 0.3500±3 | | |
| Zn$_{0.96}$Sr$_{0.04}$O | $a$=3.2467±5 | Zn/Sr | 2a | 0 | 0 | 0.8555±3 | 0.485 | 47.498±2 |
| | $c$=5.2027±1 | O | 2b | 0.3333 | 0.6667 | 0.8800±2 | | |
| Zn$_{0.94}$Sr$_{0.06}$O | $a$=3.2523±4 | Zn/Sr | 2a | 0 | 0 | 0.8690±3 | 0.464 | 47.719±1 |
| | $c$=5.2092±3 | O | 2b | 0.3333 | 0.6667 | 0.7850±4 | | |
| Zn$_{0.92}$Sr$_{0.08}$O | $a$=3.2535±2 | Zn/Sr | 2a | 0.3333 | 0.6667 | 0 | 0.409 | 47.770±2 |
| | $c$=5.2110±3 | O | 2b | 0.3333 | 0.6667 | 0.5020±1 | | |

*Table II*

| Sample | Atomic % concentration | | |
|---|---|---|---|
| | Zn | O | Sr |
| ZnO | 52.97 | 48.03 | 0 |
| Zn$_{0.92}$Sr$_{0.08}$O | 50.26 | 44.63 | 5.11 |

*Table II*

| Peak No | Undoped ZnO | | $Zn_{0.98}Sr_{0.02}O$ | | $Zn_{0.96}Sr_{0.04}O$ | | $Zn_{0.94}Sr_{0.06}O$ | | $Zn_{0.92}Sr_{0.08}O$ | |
|---|---|---|---|---|---|---|---|---|---|---|
| | Peak Position (nm) | Rel. int. (%) | Peak Position (nm) | Rel. int. (%) | Peak Position (nm) | Rel. int. (%) | Peak Position (nm) | Rel. int. (%) | Peak Position (nm) | Rel. int. (%) |
| 1 | 378 | 43 | 383 | 60 | 392 | 58 | 382 | 59 | 394 | 62 |
| 2 | 388 | 100 | 392 | 100 | 405 | 100 | 391 | 100 | 406 | 100 |
| 3 | 404 | 77 | 408 | 79 | 417 | 81 | 418 | 80 | 417 | 91 |
| 4 | 436 | 81 | 439 | 84 | 439 | 58 | 436 | 36 | 436 | 39 |
| 5 | 450 | 11 | 452 | 13 | 452 | 27 | 451 | 21 | 449 | 58 |
| 6 | 467 | 51 | 469 | 56 | 468 | 82 | 469 | 46 | 466 | 71 |
| 7 | 480 | 22 | 483 | 23 | 483 | 25 | 480 | 23 | 480 | 23 |
| 8 | 489 | 18 | 492 | 19 | 492 | 19 | 490 | 17 | 490 | 16 |
| 9 | 499 | 14 | 501 | 16 | 500 | 11 | 500 | 11 | 535 | 9 |
| 10 | 562 | 25 | 564 | 26 | 564 | 29 | 562 | 26 | 562 | 23 |

*Table III*

| Peak no | Position of the vibration bands (cm$^{-1}$) | | | | | Assignment |
|---|---|---|---|---|---|---|
| | ZnO | $Zn_{0.98}Sr_{0.02}O$ | $Zn_{0.96}Sr_{0.04}O$ | $Zn_{0.94}Sr_{0.06}O$ | $Zn_{0.92}Sr_{0.08}O$ | |
| 1 | 99.5 | 99.4 | 99.4 | 99.5 | 99.7 | $E_2^{low}$ |
| 2 | - | 277 | 277.3 | 277.8 | 278 | Defects due to $Sr^{2+}$ doping |
| 3 | 332.9 | 331.5 | 332.3 | 333.1 | 333.9 | $(E_2^{high} - E_2^{low})$ |
| 4 | 382.5 | 382.1 | 380.9 | 380.8 | 380.9 | $A_1$ (TO) |
| 5 | 438.2 | 438.8 | 438.9 | 439.3 | 438.9 | $E_2^{high}$ |
| 6 | 582.1 | 582.5 | 579.9 | 582.4 | 581.3 | LO ($A_1 + E_1$) |
| 7 | - | 663.1 | 664.8 | 663.4 | 665.7 | Defects due to $Sr^{2+}$ doping |

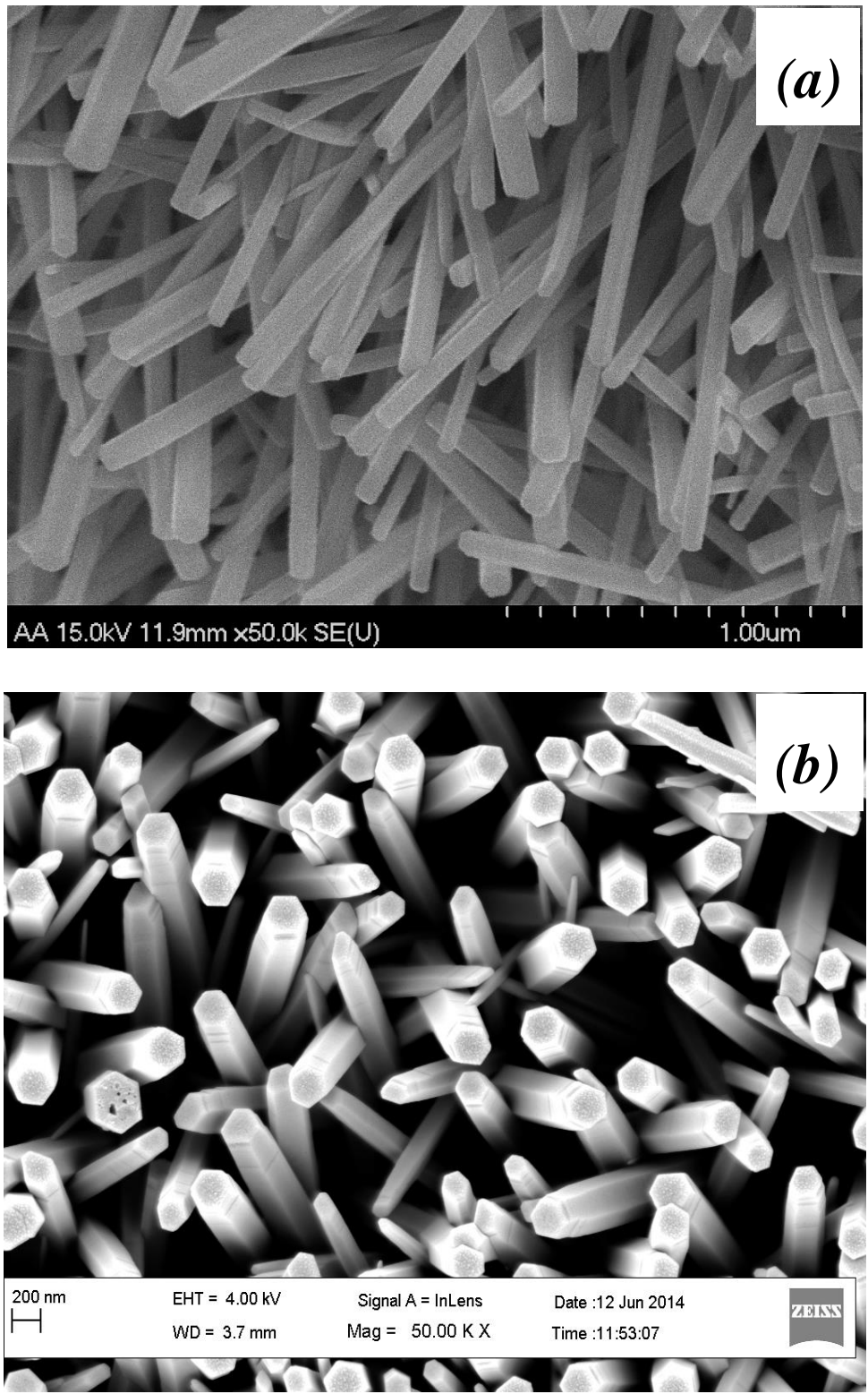

**Figure 1**

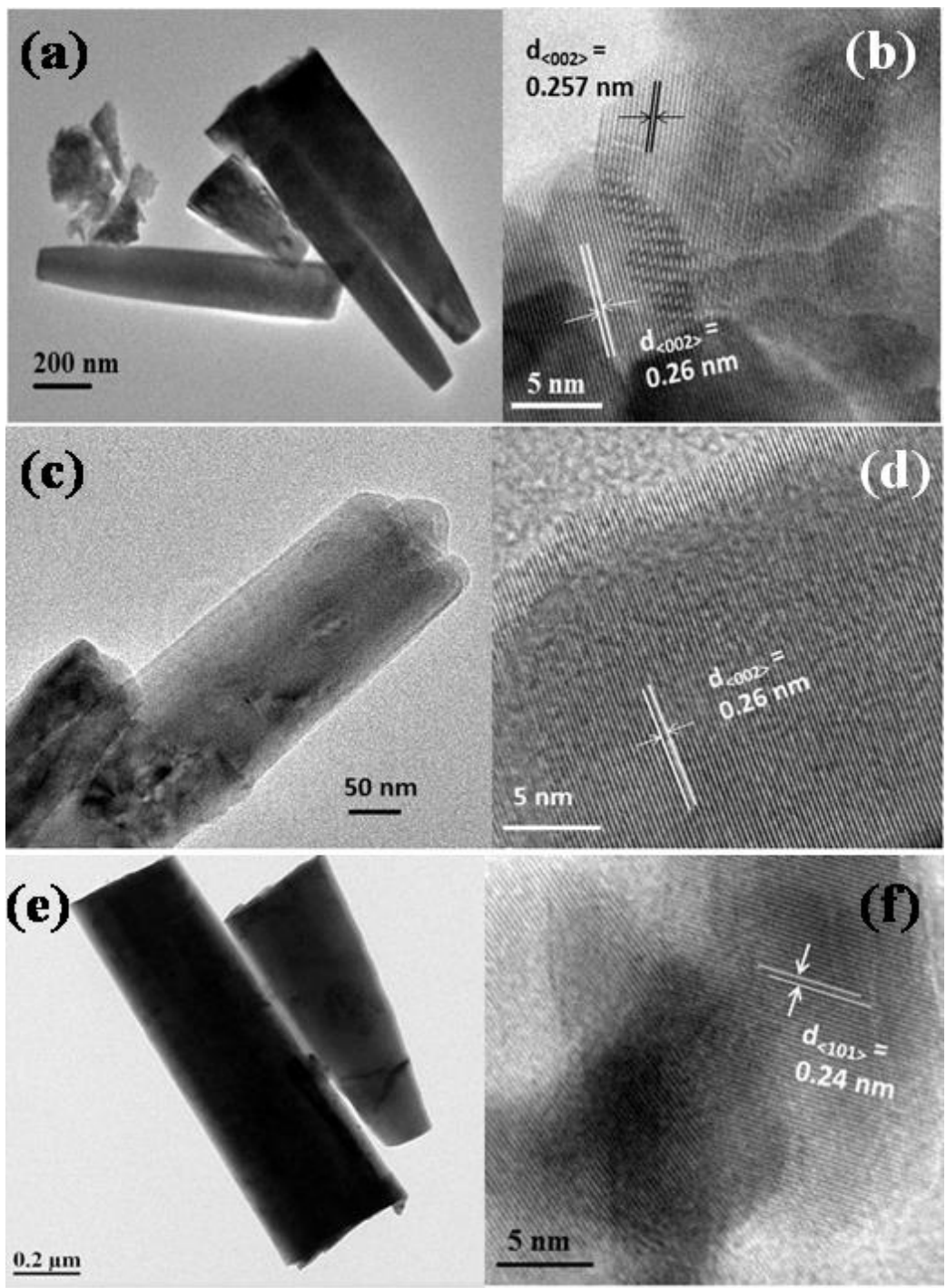

**Figure 2**

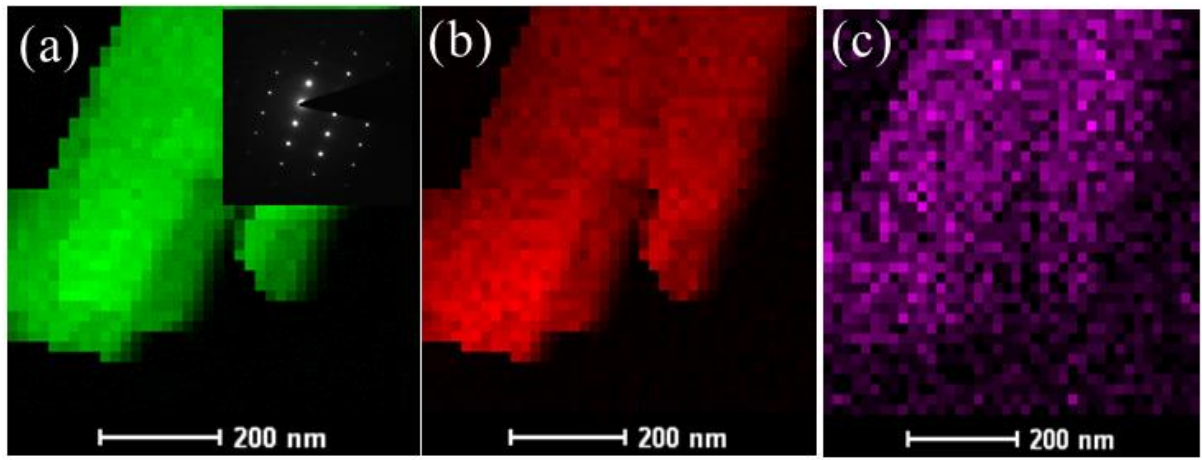

Figure 3

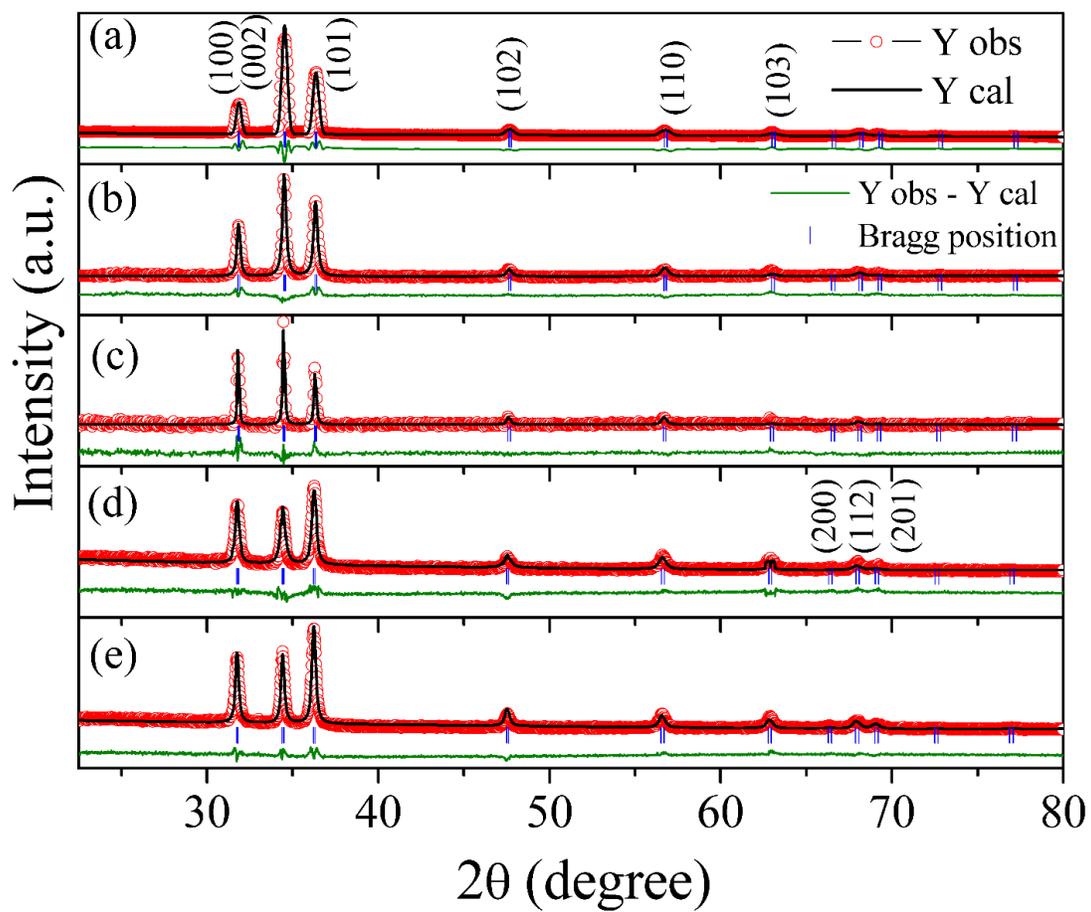

Figure 4

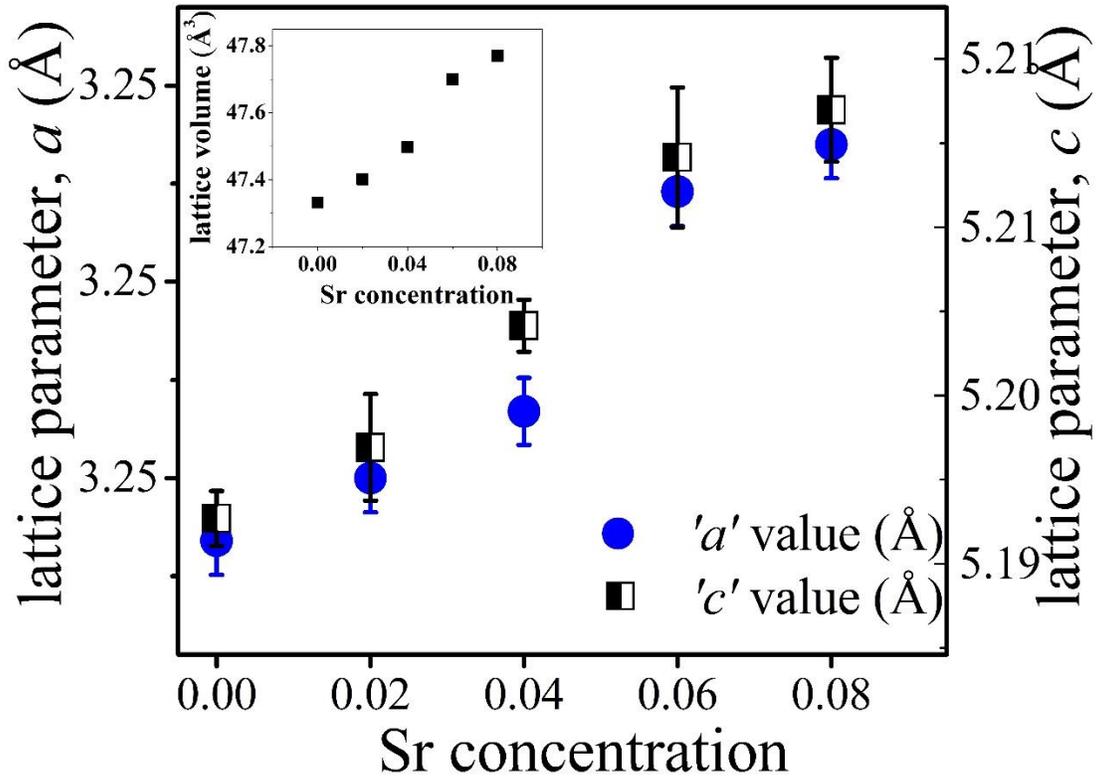

**Figure 5**

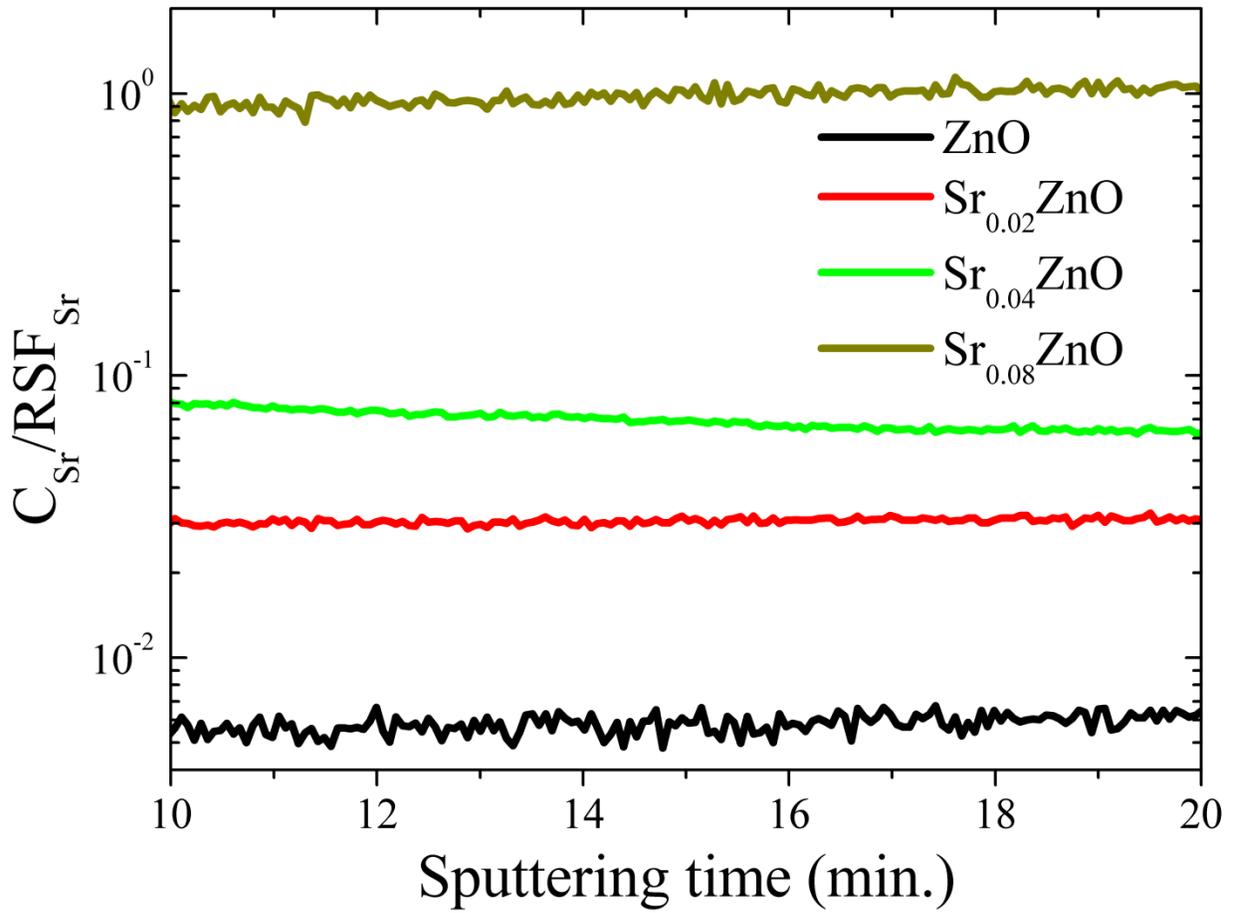

**Figure 6**

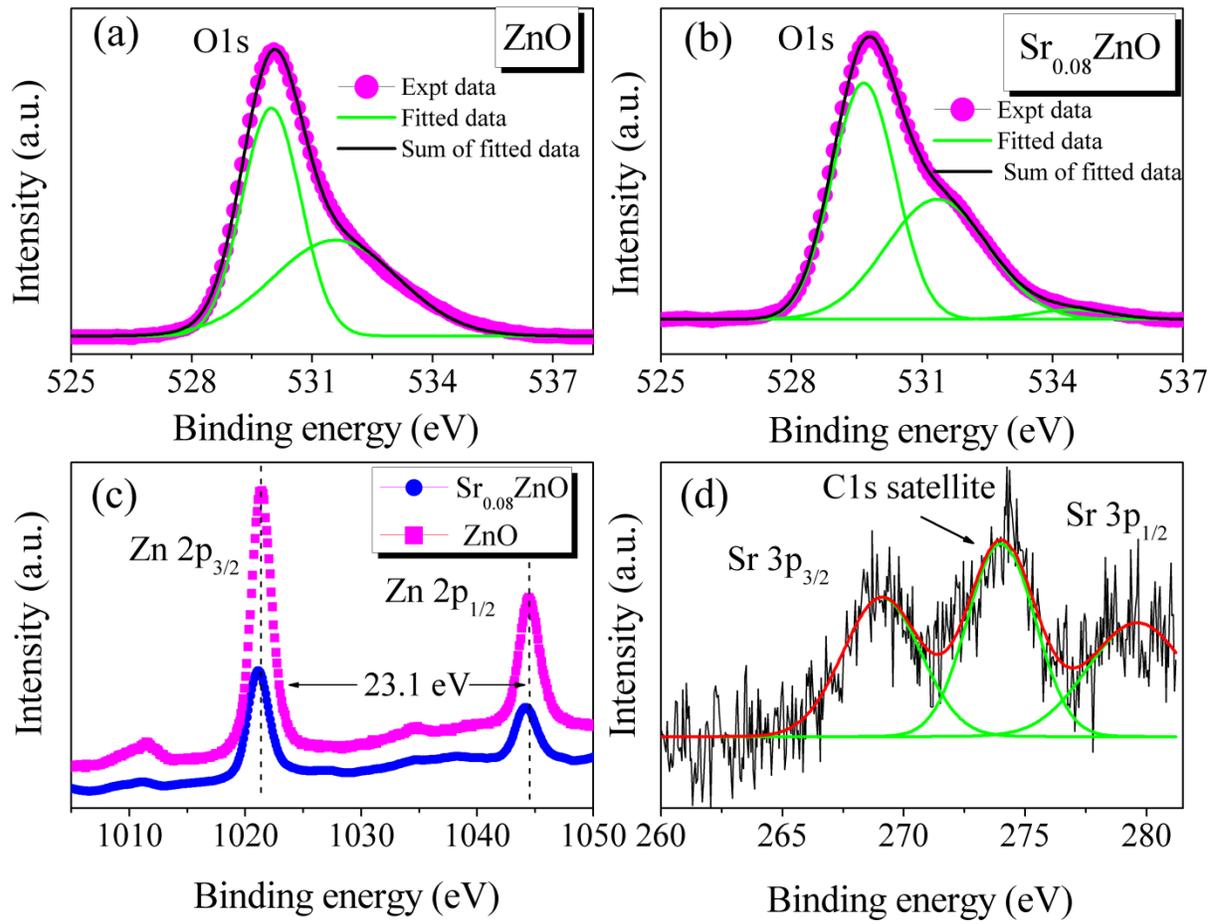

**Figure 7**

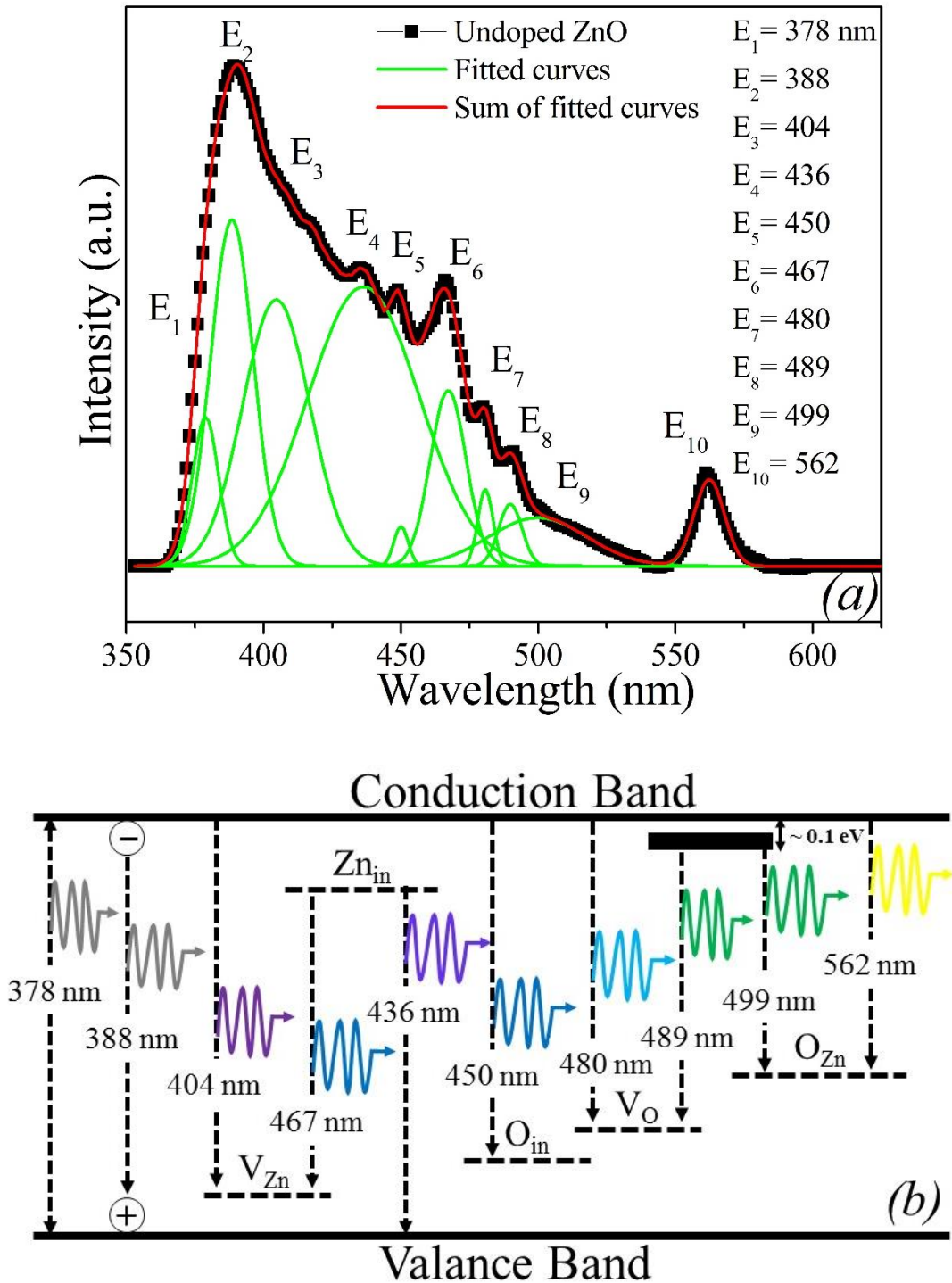

**Figure 8**

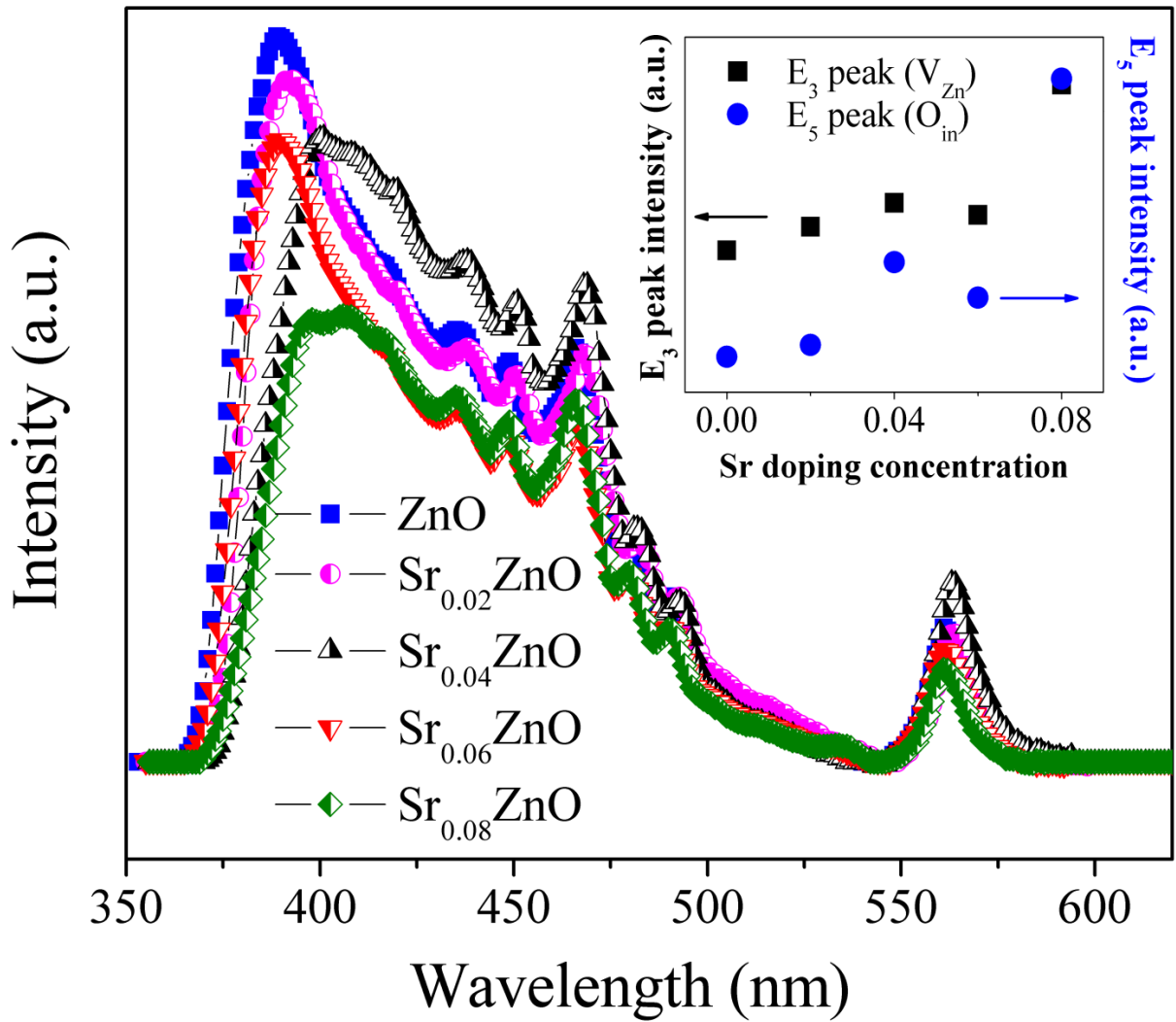

**Figure 9**

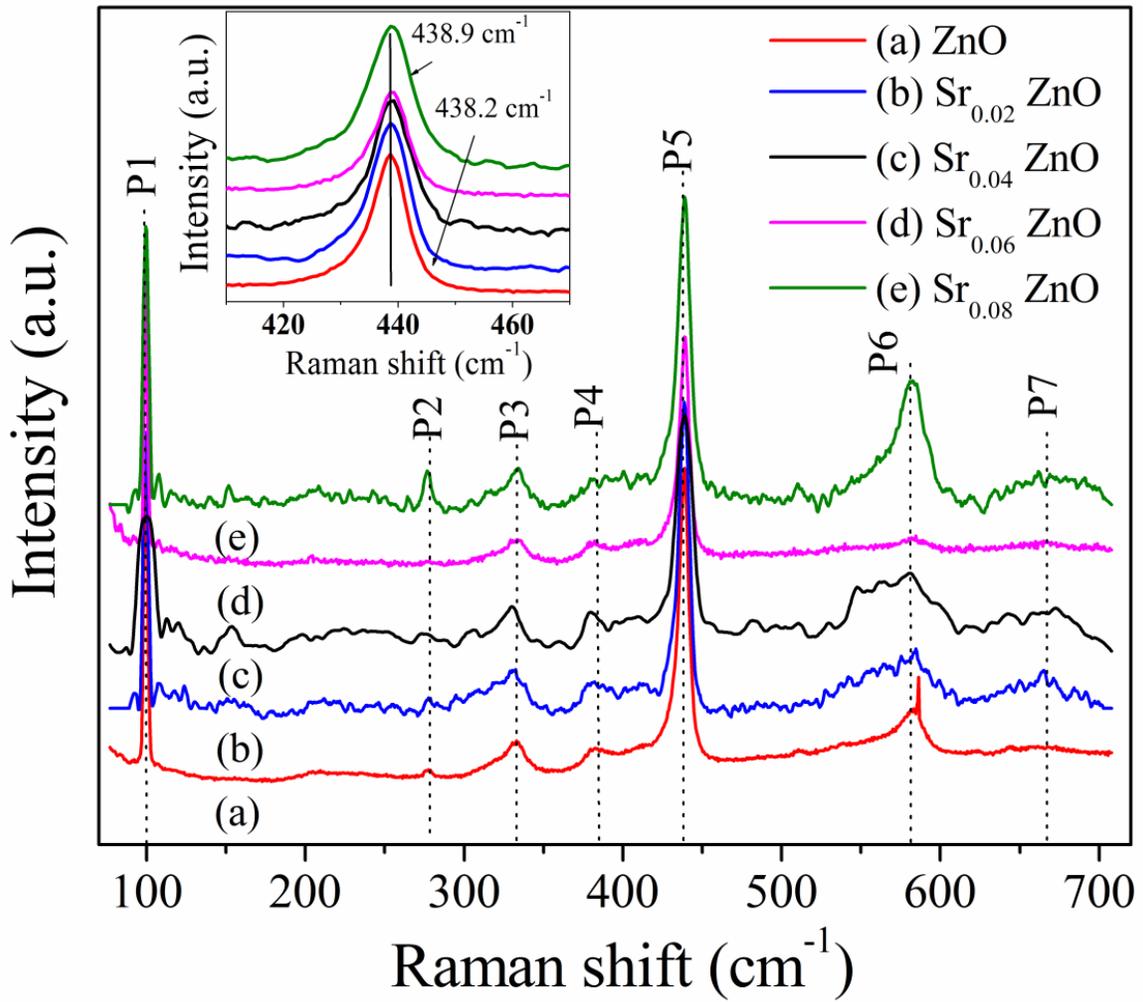

**Figure 10**

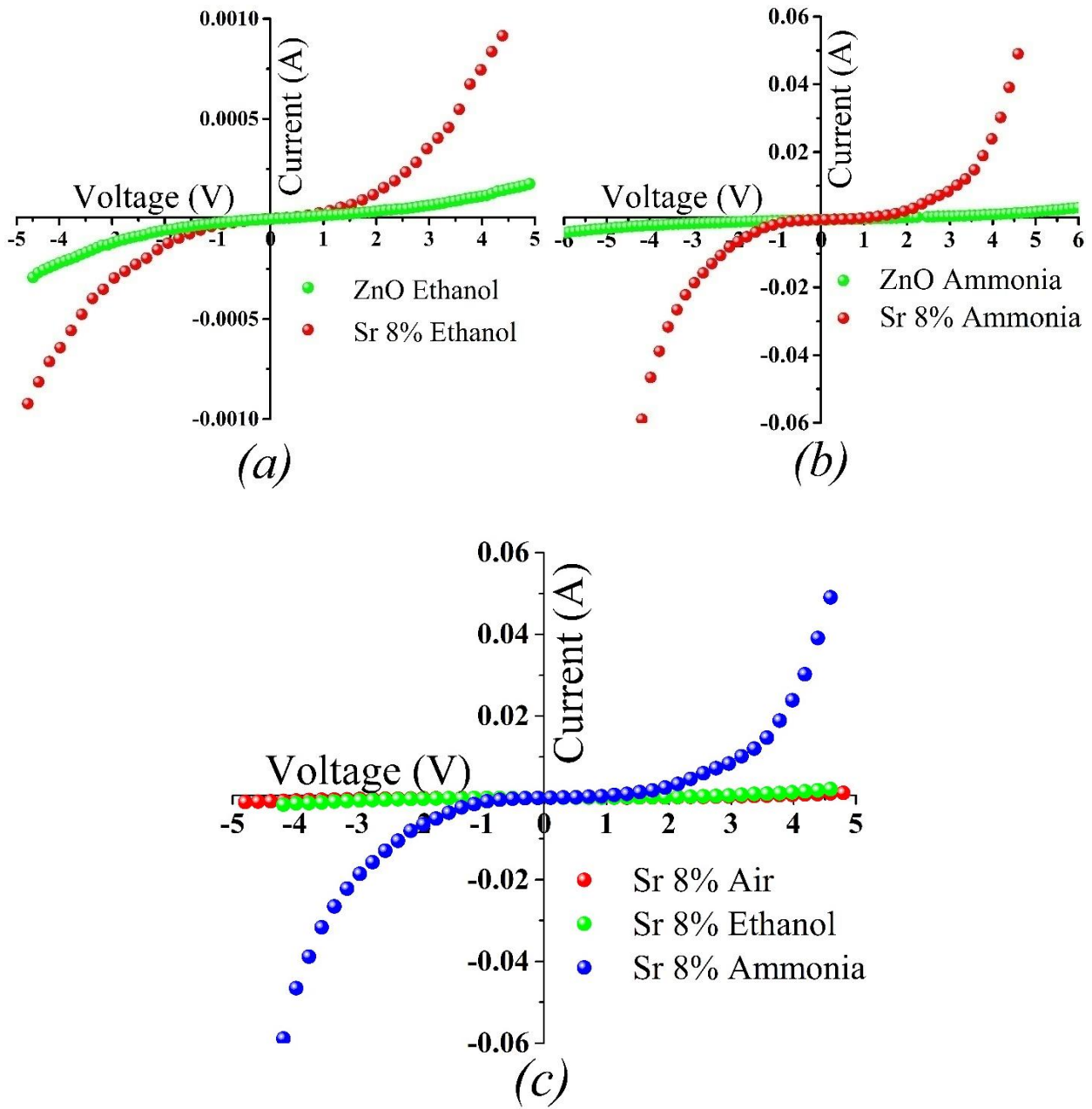

**Figure 11**

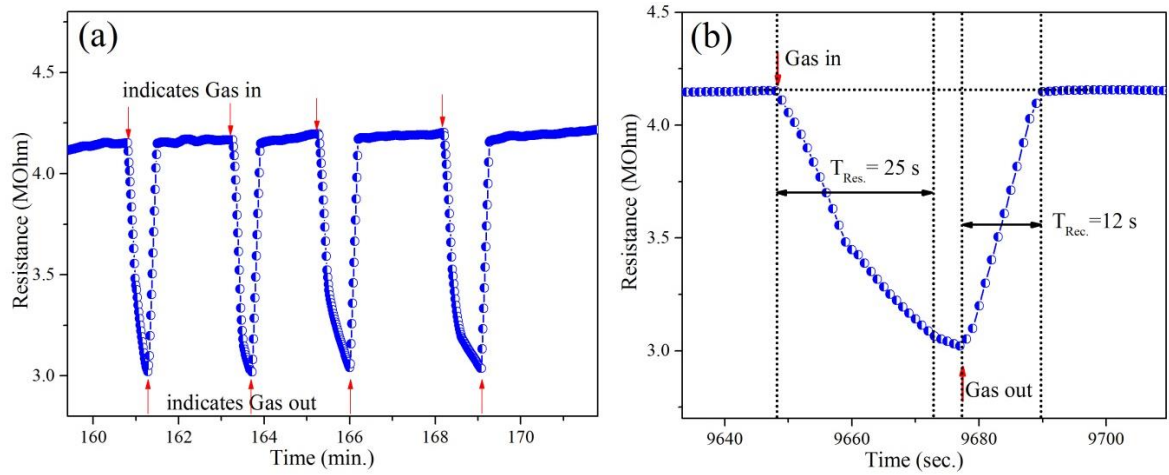

**Figure 12**